\documentclass[
twocolumn,
 amsmath,amssymb,
 aps,
superscriptaddress]{revtex4-2}

\usepackage{xcolor}
\usepackage{graphicx}
\usepackage{dcolumn}
\usepackage{bm}
\usepackage{float}
\usepackage{upgreek}
\usepackage{hyperref}
\usepackage[capitalize]{cleveref}

\newcommand{\tpm}[2]{#1#2}
\newcommand{\revision}[1]{#1}

\begin{document}

\preprint{APS/123-QED}

\title{Time-bin entanglement in the deterministic generation \\ of linear photonic cluster states}

\author{David Bauch}
\affiliation{Department of Physics and Center for Optoelectronics and Photonics Paderborn (CeOPP), Paderborn University, Warburger Strasse 100, 33098 Paderborn, Germany}%
\affiliation{Institute for Photonic Quantum Systems (PhoQS), Paderborn University, 33098 Paderborn, Germany}

\author{Nikolas Köcher}
\affiliation{Department of Physics and Center for Optoelectronics and Photonics Paderborn (CeOPP), Paderborn University, Warburger Strasse 100, 33098 Paderborn, Germany}%
\affiliation{Institute for Photonic Quantum Systems (PhoQS), Paderborn University, 33098 Paderborn, Germany}

\author{Nils Heinisch}
\affiliation{Department of Physics and Center for Optoelectronics and Photonics Paderborn (CeOPP), Paderborn University, Warburger Strasse 100, 33098 Paderborn, Germany}%
\affiliation{Institute for Photonic Quantum Systems (PhoQS), Paderborn University, 33098 Paderborn, Germany}

\author{Stefan Schumacher}
\affiliation{Department of Physics and Center for Optoelectronics and Photonics Paderborn (CeOPP), Paderborn University, Warburger Strasse 100, 33098 Paderborn, Germany}%
\affiliation{Institute for Photonic Quantum Systems (PhoQS), Paderborn University, 33098 Paderborn, Germany}
\affiliation{Wyant College of Optical Sciences, University of Arizona, Tucson, AZ 85721, USA}%

\date{\today}
             
\begin{abstract}
\revision{We theoretically investigate strategies for the  deterministic creation of trains of time-bin entangled photons using an individual quantum emitter described by a $\Lambda$-type electronic system. We explicitly demonstrate the theoretical generation of linear cluster states with substantial numbers of entangled photonic qubits in full microscopic numerical simulations}. The underlying scheme is based on the manipulation of ground state coherences through precise optical driving. One important finding is that the most easily accessible quality metrics, the achievable rotation fidelities, fall short in assessing the actual quantum correlations of the emitted photons in the face of losses. To address this, we explicitly calculate stabilizer generator expectation values as a superior gauge for the quantum properties of the \revision{generated} many-photon state. \revision{With widespread applicability also to other emitter and excitation-emission schemes, our work lays the conceptual foundations for an in-depth practical analysis of time-bin entanglement based on full numerical simulations with predictive capabilities for realistic systems and setups including losses and imperfections.} The specific results shown in the present work illustrate that with controlled minimization of losses and realistic system parameters for quantum-dot type systems, useful linear cluster states of significant lengths can be generated in the calculations, discussing the possibility of scalability for quantum information processing endeavors.
\end{abstract}

\maketitle

Quantum information theory and processing are poised to fundamentally transform computational paradigms, surpassing the limitations of classical systems \cite{kimble2008quantum}. Central to this quantum leap is the development of reliable and deterministic photon emitters \cite{Ding2016OnDemandSinglePhotons,schweickert2018demand,Huber2008StrainEntangledOnDemand,liu2019solid,gong2010linear,jennewein2011single,briegel2001persistent,jonas2022nonlinear,praschan2022pulse,heinisch2024swing,karli2022super,bracht2021swing,kuhlmann2015transform,chow2013physics,michler2000quantum,heindel2017bright}, which are essential for the encoding, transmission, and manipulation of quantum information in measurement-based quantum computing \cite{raussendorf2001one,claes2023tailored} and quantum communication \cite{vaziri2002experimental,bennett2014quantum,pan2012multiphoton,walther2005experimental,chen2007experimental,bozzio2022enhancing,vajner2022quantum,lodahl2017quantum}. Multi-photon states with significant quantum correlations or entanglement are a critical ingredient for numerous quantum protocols \cite{pan2012multiphoton,jennewein2000quantum,meng2023photonic} and the quest for their robust generation forms a challenging field of research \cite{schwartz2016deterministic,zhang2006experimental}. Besides polarization \cite{seidelmann2022two,jennewein2000quantum} and other degrees of freedom \cite{huber2018semiconductor,juska2013towards,bauch2021ultrafast,bauch2023demand}, quantum correlations can be encoded in the form of time-bin entanglement \cite{shi2021fibre,jayakumar2014time,nutz2017proposal,lee2019quantum,aumann2022demonstration} of photonic qubits \cite{jayakumar2014time,koutny2023deep}, which holds great potential for applications in quantum teleportation and quantum key distribution \cite{avoidingLeakageAndErrors,nutz2017proposal,heindel2012quantum}.

In the present work, we delve into the generation of time-bin entangled photonic states with a $\Lambda$-type electronic system. The system enables the individual excitation of the system's ground states to corresponding excited states to produce photons, while the coherent superposition of the ground states can be controlled through optical driving of the lambda transitions. This approach is quite broadly applicable to systems exhibiting lambda configurations, including, but not limited to, color centers \cite{avoidingLeakageAndErrors,adambukulam2024coherent,santori2006coherent,pieplow2023deterministic}, stacked quantum dots \cite{gimeno2019deterministic,economou2010optically}, and quantum dot molecules \cite{vezvaee2022deterministic,economou2012scalable}, commonly using trion-based spin-hole qubits \cite{adelsberger2022hole,avoidingLeakageAndErrors}. Entanglement between individual time bins is achieved by alternating photon emission and ground state rotations, which in theory allows for the generation of complex cluster and graph states \cite{raissi2022deterministic,lu2007experimental}. In the present work we go beyond the development of a schematic protocol and actually implement, benchmark, and optimize a specific excitation scheme. We explicitly assess quantum correlations of the emitted photons via time-shifted second- and third-order correlation functions in advanced numerical simulations. This rigorous analysis allows us to estimate the lower bounds of time-bin entanglement by calculating expectation values of stabilizer generators \cite{nutz2017proposal}. We also explore photon indistinguishability, uncovering the capability to generate sequences of photonic qubits with robust time-bin entanglement and high indistiguishability, even from lossy quantum emitters \cite{cogan2023deterministic,su2024continuous}. In comparison with our analysis of photon correlation functions, we demonstrate that gate fidelities are generally not a sufficient measure of system performance. With losses, even for almost perfect gate fidelities and excitation-emission protocol implementation, quantum correlations in the generated photon emission can in fact be very low. Our microscopic theory includes dynamical light-field induced shifts and different types of losses, which leads to predictive capabilities for realistic systems and structures. The present work not only advances our understanding of photonic entanglement generation but also reinforces the foundational techniques necessary for the next generation of quantum information processing.

\begin{figure}[t]
    \centering
    \includegraphics[width=0.9\linewidth]{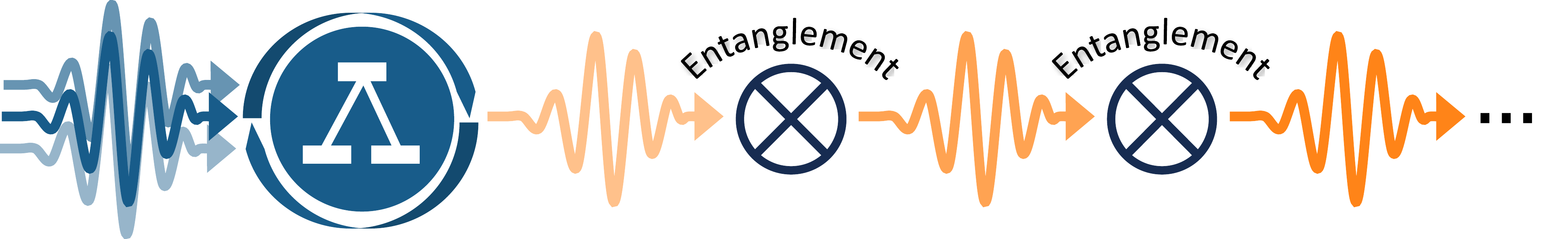}
    \caption{\textbf{Schematic of a $\Lambda$-type photonic quantum emitter.} The emitter is periodically driven by a sequence of laser pulses that introduce excitations and perform entangling operations. Emission from the emitter's excited states results in generation of trains of quantum correlated photons. Depending on the specific excitation-emission protocol, entanglement is manifested in polarization or temporal degrees of freedom of the generated many-photon state.}
    \label{fig:emission}
\end{figure}

\section{$\Lambda$-Type Emitter Structure}

The design of the emitter structure is pivotal, as it must exhibit a $\Lambda$-shaped energy configuration of the underlying electronic states. Such a configuration enables us to achieve dual objectives: firstly, to excite individual electronic states, leading to the emission of photons via radiative decay; and secondly, to infuse the emitted photons with the desired quantum phase information, a direct consequence of the coherences between the states within the $\Lambda$ system, which is schematically depicted in \cref{fig:emission}. The capability to \revision{instill coherence of the quantum state} is essential for the generation of photon entanglement. The temporal entanglement of photonic qubits, as produced by this structure, plays a foundational role in a wide array of quantum protocols, and thus constitutes a key mechanism for advancing the field of quantum information science. For example, specific types of quantum dot molecules are known to exhibit such energy structure, making them a prime solid-state candidate for realizing the $\Lambda$-type configurations required.

\begin{figure}[!t]
    \centering
    \includegraphics[width=1\linewidth]{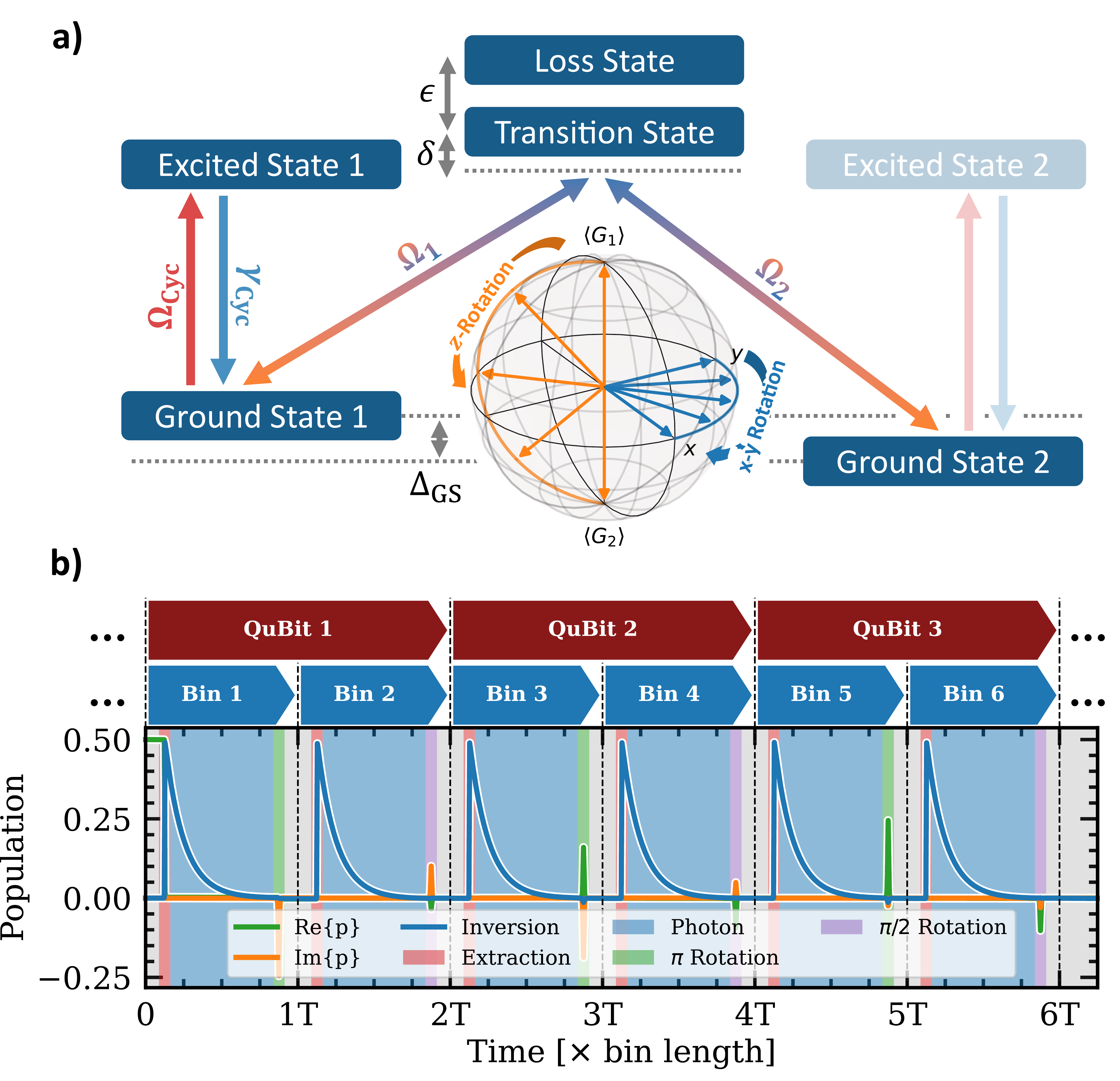}
    \caption{\textbf{Energy-level structure and emission scheme}. \textbf{a)} The $\Lambda$ system depicted, which may for example be realized in quantum dot molecule systems, features two ground states which are separated by a finite energy splitting $\Delta_\text{GS} = E_{G_2}-E_{G_1}$. Both ground states $G_1$ and $G_2$ couple to excited states, which can emit a single photon through radiative decay with rate $\gamma_\text{Cyc}$; for simplicity only the first excited state is used for photon emission here. Consequently, the second excited state is depicted in a faded manner in the illustration. Both excited states can be driven using a laser pulse $\Omega_\text{Cyc}$. The transition states are driven by two additional laser pulses $\Omega_{1,2}$ and are split energetically by $\epsilon$. The desired rotations in the coherent superposition of the two ground states can be achieved by adjusting the transition pulse detunings using $\delta > 0$. \textbf{b)} Emission sequence simulated for \revision{a combined total of three} photons as an example; each photon is emitted into a time bin, pairs of subsequent early and late time bins form a photonic qubit. Quantum correlations are defined between the states of these qubit. Pulses are hyperbolic sechant shaped, with lengths of 16ps for rotation and 5ps for extraction pulses. Radiative decay is present for the cycling transitions in the scenario shown here, losses for other states and transitions are included further below as discussed in the text. Inset: Bloch sphere illustrating rotations in the x, y plane and along the z axis.}
    \label{fig:temp_system}
\end{figure}

The specific $\Lambda$-type system investigated here is characterized by its two distinct ground states, the excited states and the transition and loss states. Regardless of the physical implementation, the ground states will exhibit some degree of spectral splitting $\Delta_\text{GS}$. We will call the ground state that is spectrally closer to the transition states \textit{first ground state}, and the corresponding other \textit{second ground state}. The ground state splitting can be achieved e.g. by using stacked quantum dots. \revision{Small alignment mismatches will then result in mixing of the states, enabling rotations through the transition states.} Each of the ground state transitions can be individually driven by differently polarized laser pulses $\Omega_\text{Cyc}$, offering high excitation fidelity of each specific cycling transition. The excited states will exhibit radiative decay; here we assume a radiative lifetime of $1/\gamma_\text{Cyc} = 0.5$ns, which results in \revision{an almost} full recombination of the excited states within about $3$ns. This is depicted in \cref{fig:temp_system}~b) and lies well within the realistic range for commonly manufactured quantum dots \cite{hopfmann2021maximally,birowosuto2012fast,bauch2023demand,bauch2021ultrafast}. For the sake of simplicity in our analysis, we only consider photon emission from the excited state associated with the first ground state, even though our numerical model also includes the potential for emissions linked to the second ground state's excited state. 
Both ground states couple to several transition states. In this case, the transition states are limited to a single target transition and a single unwanted transition, slightly reducing rotation fidelities. Moreover, both ground states couple to a superposition of available transition states. These transitions are driven by the optical pulses $\Omega_{1/2}$, targeting both the target and the unwanted state. We set the excited state energy as well as the transition state energy to a fixed energy $\hbar\omega = 1$eV, which is well within the values for commonly used QD materials \cite{birowosuto2012fast,bauch2023demand,liu2019solid}. We note that the system performs analogously also for different energies. We further adopt viable system parameters from previous work \cite{vezvaee2022deterministic}, setting the target-unwanted splitting to $\epsilon = 500\upmu$eV and fixing the ground state splitting to $\Delta_\text{GS} = 200\upmu$eV. \revision{For definition of the system in form of its Hamiltonian see Appendix~A.} Setting both values to zero reduces the transition system to a simple degenerate $\Lambda$ system, avoiding the need to phase tune the rotation pulses. By fixing parameters such as the excited and transition state energies, along with specific energy splittings, we intentionally limit the variables in our study. This approach streamlines our focus on essential factors that significantly impact the system's ability to generate entangled streams of photons (avoiding a full optimization of the system, which could be a part of future work). For the rotation pulses, we use hyperbolic secant shaped envelopes defined by 
\begin{align}
    \Omega(t) = \alpha\sigma\text{sech}\left[ \sigma(t-t_0) \right] \cdot \text{exp}\left[ -i(\omega (t-t_0) + \vartheta) \right] ~,
\end{align}
with pulse amplitude $\alpha$, pulse bandwidth $\sigma$, center $t_0$ and pulse frequency $\omega$. Through the bandwidth, we set the pulse length to a fixed $\tau_\text{Transition} = \frac{1}{\sigma} = 16$ps for the transition pulses and $\tau_\text{Cyc} = 5$ps for the excitation pulses. The pulse width translates into a pulse bandwidth of $\approx 41\upmu$eV$/\hbar$, which, in previous work \cite{avoidingLeakageAndErrors}, proved to be around the ideal bandwidth for the given energy configuration. To achieve a desired rotation of the ground state superposition by angle $\Theta$, we set the pulse amplitude of the total driving force of the system to $\alpha_1^2 + \alpha_2^2  = 1$ and use a finite non zero detuning of the pulse energy. Prior work has shown that equal coupling to both states is most effective in achieving high fidelity rotations of the ground state coherences \cite{avoidingLeakageAndErrors}. Therefore, we limit our investigation to cases of equal coupling where both pulses couple equally to both of the transition states. This means, that both the unwanted and the target state pulses use the same amplitude and phase, while the phases between pulses driving transitions for different ground states can differ. However, we note that a generalization to arbitrary coupling rates is easily achievable and merely increases the size of the parameter space to be analyzed.

Accounting for the ground state splitting, the resonant pulse frequencies are $\omega$ and $\omega - \omega_\text{GS}$, respectively.
Since the coupling ratio is not an explicit part of our investigation, we standardize the pulse coupling for each transition to a value of one, acknowledging that selecting other couplings high-fidelity rotations can also be achieved, albeit at different detuning values.

We simulate the temporal dynamics of the system using the von Neuman equation $\frac{\partial \rho}{\partial t} = -\frac{i}{\hbar}[H,\rho] + \sum_{\hat{O}}\frac{\gamma_O}{2}\left( 2\hat{O}\rho\hat{O}^\dagger - \hat{O}^\dagger\hat{O}\rho - \rho\hat{O}^\dagger\hat{O}\right)$ with Lindblad-type losses, where $\hat{O}$ denotes any of the available system state operators and $\gamma_O$ is the corresponding loss rate. \revision{The density matrix $\rho$ in matrix representation is propagated in time using explicit numerical integration methods and using the Hamiltonian in \cref{hamiltonian}}. We distinguish between the following decays or loss channels: the radiative decay of the cycling transition ($\gamma_\text{Cyc}$), the decay of both the target and unwanted state population ($\gamma^\text{R}_\text{T,U}$), and the spin dephasing rate of the ground states ($\gamma^\text{D}_\text{T,U}$). Unless specified otherwise, we assume that both $\gamma^\text{D}_\text{T,U}$ and $\gamma^\text{R}_\text{T,U}$ are set to zero. 

\begin{figure}[!t]
    \centering
    \includegraphics[width=1\linewidth]{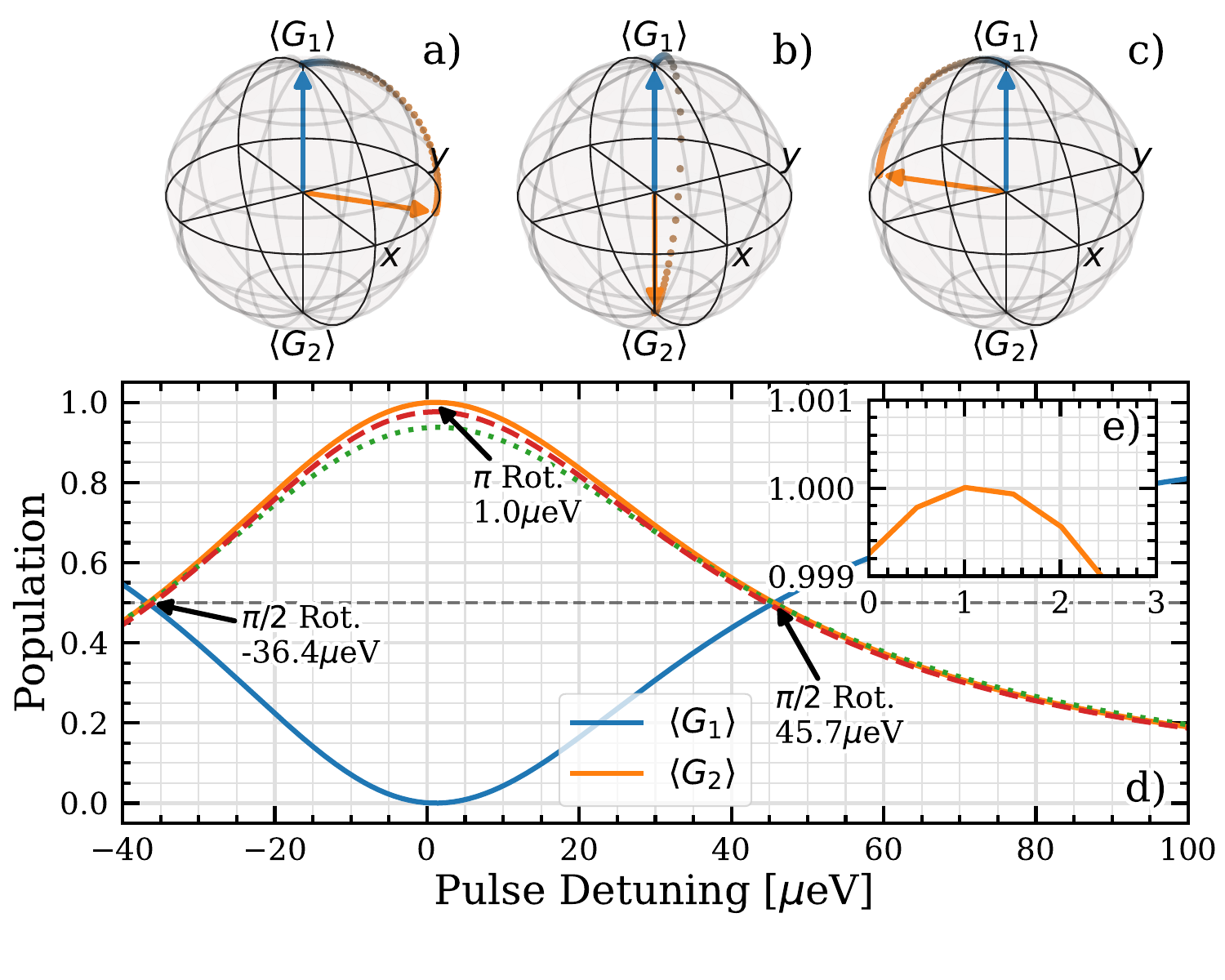}
    \caption{\textbf{Sensitivity to detuning of rotation pulses}. Panels \textbf{a-c)} show Bloch-sphere representations of the rotations annotated in \textbf{d)} from left to right. The blue arrow indicates the initial state, the orange arrow indicates the final state. The spheres show rotations for the lossless case with normalized state vectors. Panel \textbf{d)} shows expectation values for the first and second ground state population for varying pulse detuning of the rotation pulses; the pulse amplitudes are fixed to $\Omega_\text{1,2} = \frac{\pi}{\sqrt{2}}$. For a detuning of zero, both pulses are resonant with their respective transitions. The inset \textbf{e)} depicts a zoom on the $\pi$ rotation. Dotted lines show rotations with exaggerated relaxation of the target and unwanted states ($\gamma^\text{R}_\text{T,U}=2\gamma_\text{Cyc}$, green dotted line). Dashed lines show rotations with exaggerated spin dephasing of the ground states ($\gamma^\text{D}_\text{T,U}=\gamma_\text{Cyc}$, red dashed line).}
    \label{fig:find_rotation}
\end{figure}

Our investigation begins with finding suitable rotation pulses. In \cref{fig:find_rotation} we vary the pulse detuning from $0$ to $100\upmu$eV. We find that for our choice of parameters, a ground state rotation of $\Theta = 1\pi$ is achieved for a detuning of $\delta = \hbar\omega - \hbar\omega_1 - \Delta_\text{GS} \approx 1\upmu$eV. A rotation of $\Theta = 0.5\pi$ is achieved for a detuning of either $\delta \approx 45.7\upmu$eV or $\delta \approx 36\upmu$eV. The latter agrees well with the analytical expression 
\begin{align}
    \delta = \frac{1}{2}\left( \epsilon \pm \sqrt{\epsilon^2 + 4\epsilon\sigma\text{cot}\left(\Theta/2\right)-4\sigma^2} \right) \label{eqn:analytical_formula}
\end{align}
for the captured ground state phase \cite{avoidingLeakageAndErrors} after a given transition pulse for a target angle of either $\Theta = 1\pi$ or $\Theta = 0.5\pi$. Following the analytical formula, \cref{eqn:analytical_formula}, to calculate these values, we find detunings for $\pi$ of $\Delta = 3.5\upmu$eV and for $\pi/2$ of $\Delta = -35\upmu $eV. This is true even though \cref{eqn:analytical_formula} is not fully valid for the system in place because it relies on slightly different couplings and does not incorporate the ground state splitting or any light-field induced Stark shifts of the energies.

\begin{figure}[!t]
    \centering
    \includegraphics[width=1\linewidth]{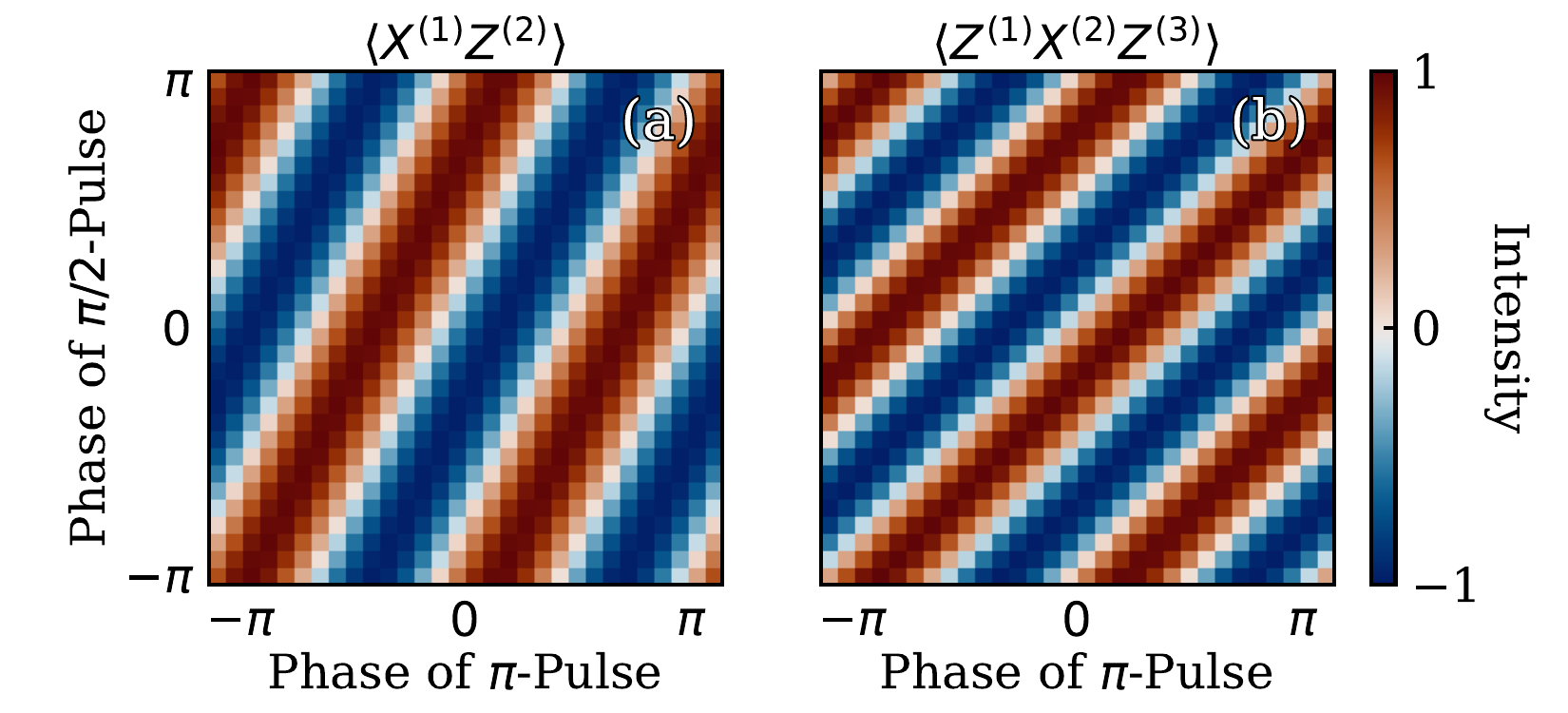}
    \caption{\textbf{Phase dependence of stabilizer generator expectation values}. Shown are real parts of the first two stabilizer generator expectation values for the emission depicted in \cref{fig:temp_system}~b). The rotation pulses are optimized to $>99\%$ gate fidelity; we vary the phase $\vartheta$ of the $\pi$ and $\pi/2$ rotations, respectively. The resulting average stabilizer magnitudes are $|\overline{\Phi Z}| = 0.96$ with a maximum $|\Phi Z| = 0.99$ and $|\overline{Z\Phi Z}| = 0.95$, with a maximum $|Z\Phi Z| = 0.99$.}
    \label{fig:phase_sweep}
\end{figure}

We find in \cref{fig:phase_sweep}, that the choice of detuning is arbitrary for the resulting phase dependency, as we can freely chose other pulse phases to compensate for different choices of pulse detunings. We find that high fidelity $\pi/2$ rotations are always possible, while high fidelity $\pi$ rotations are limited by system losses. 

It is also noteworthy that small dephasing rates of the transition states have a marginal negative impact on the rotation fidelities. Conversely, even relatively low decay rates of the target and unwanted states considerably diminish the rotation fidelities. Previous research has corroborated these findings, though it did not assess the translation of rotation fidelities into quantum correlations in the emitted photonic states, an analysis we undertake in the following with the present study.

\section{Time-Bin Emission\label{sec:timebinphotons}}
Time bins serve as a fundamental concept in our study, essentially acting as discrete temporal slots into which photons can be emitted. These bins are critical for encoding quantum information, where the presence or absence of a photon in a specific time slot represents a logical 1 or 0. We define \textit{early} and \textit{late} time bins to create a pair that forms a single \textit{photonic qubit}. The early time bin, denoted as $|\text{early}\rangle = |1_{\tau=i}0_{\tau=i+1}\rangle = |0\rangle$, corresponds to the emission of a photon in the initial time slot, whereas the late time bin, $|\text{late}\rangle = |0_{\tau=i}1_{\tau=i+1}\rangle = |1\rangle$, represents photon emission in the subsequent time slot. This binary representation allows us to encode and manipulate quantum information across a sequence of time bins, effectively creating a framework for generating complex quantum states. It is crucial for the photons of these time bins to demonstrate particular phase relationships to ensure the quantum correlations are high. 
We use the notation $R_{\vartheta}(\Theta)$ to denote rotations in the Bloch sphere of the two ground states, with $\vartheta$ being the angle in the x-y-plane and $\Theta$ being the angle in the z direction.
As an explicit example, in the present study we limit our investigation to the following excitation-emission scheme to generate a linear cluster state \cite{vezvaee2022deterministic,istrati2020sequential,vallone2010six}:
\begin{itemize}
    \item Start or prepare the system in $|\Psi\rangle = \frac{1}{\sqrt{2}}\left(|\text{G}_1\rangle + |\text{G}_2\rangle\right)$
    \item Step 1: Drive $\Omega_\text{Cyc}$ using a $\pi$-pulse, followed by emission of first photon
    \item Step 2: Perform $R_\vartheta(\pi)$ through the $\Lambda$-System
    \item Step 3: Drive $\Omega_\text{Cyc}$ using a $\pi$-pulse, followed by emission of second photon
    \item Step 4: Perform $R_\vartheta(\pi/2)$ through the $\Lambda$-System
    \item Repeat steps 1-4 N times
    \item Perform a final Z projection to decouple the emitter\,.
\end{itemize}
Steps 1-4 are repeated N times to create an N qubit linear cluster states. The final step, involving the removal of the emitter from the cluster state, is not always necessary and depends on the specific desired state \cite{raissi2022deterministic}. The photon emission occurs through spontaneous radiative emission from the corresponding excited state. The x-y angle $\vartheta$ is tunable by selecting appropriate pulse phases, while $\Theta$ can be selected varying the pulse amplitudes or detunings. We consistently select a bin length that accurately encompasses the emitted photons, in accordance with $\gamma_\text{Cyc}$. For a radiative decay rate of $\gamma_\text{Cyc} = 1.2\upmu$eV ($\approx 2$GHz), we find that a time bin length of $T = 4$ns suits the emission scheme well as it incorporates both the emission process as well as a time buffer to apply the rotation pulses. Additionally, we choose rotation pulses of the highest fidelity, ensuring that the pulse bandwidth is appropriately matched to the selected splittings. In \cref{fig:temp_system} b), the emission of up to \revision{three} photons is illustrated; it is important to note though that the system or protocol are not constrained to this particular number of photons.

We observe that the electronic coherences disappear immediately upon the application of the $\pi$-extraction pulse through the cycling transition, provided the system is in an equal superposition of both ground states. \revision{Further $\Theta = \pi$ or $\Theta = \pi/2$ rotations through the $\Lambda$-system are then not visible withing the electronic coherences and, in an ideal case, leave the ground state population unchanged}. For this reason, these pulses are referred to as \textit{transitionless} \cite{avoidingLeakageAndErrors}. This phenomenon is by design, as it facilitates the transfer of coherences from the electronic states onto the emitted photons.

Although our research focuses on the previously described scheme, it is feasible to adopt alternate excitation-emission protocols by adjusting the rotation pulses and with expanding the present $\Lambda$-system accordingly. Such modifications can then lead to the generation of higher-order cluster \cite{briegel2001persistent,wunderlich2009two}, Greenberger-Horne-Zeilinger (GHZ) \cite{toth2005entanglement,zheng2001one,jung2008greenberger,meng2023deterministic,tiurev2022high} or graph states \cite{adcock2019programmable}, expanding the scope of potential quantum states achievable. Moreover, while our focus is on time-bin entanglement, the adoption of polarization entanglement is conceivable \cite{lu2007experimental}, albeit with the trade-off of potentially constraining the diversity of quantum states that can be generated. Furthermore, the approach to measuring and characterizing cluster states significantly influences their classification. Specifically, the "horseshoe" configuration presents an interesting case within the broader category of 1D linear cluster states. This distinction highlights the complexity in cluster state classifications and emphasizes the nuanced understanding required to fully leverage their potential in quantum information science \cite{vallone2010six}.

\section{Stabilizer Generator Expectation Values}
For a full quantum state tomography or detailed analysis of photon entanglement in a given $N$-photon quantum state, one might consider computing the $N$-photon density matrix directly. From here, protocol fidelities can be calculated \cite{tiurev2021fidelity}. However, in the microscopic simulations, this approach becomes computationally intractable, typically already for $N>3$, due to the exponential increase in space dimensionality and requirement for the explicit evaluation of N-th order photon correlation functions $\mathcal{G}^{(N)}$, \revision{also compare with \cite{bracht2024theory} for details on $\mathcal{G}^{(1)}$ and $\mathcal{G}^{(2)}$}. 
To keep the computational effort manageable, here we utilize stabilizer generator expectation values \cite{toth2005entanglement} as our primary metric for evaluating quantum correlations in the generated $N$-photon quantum states of interest. These values allow a localized investigation of entanglement by probing correlations in a segmented fashion, thereby significantly reducing computational demands. This is accomplished by evaluating correlation functions $\mathcal{G}^{(3)}$ only up to third-order, resulting in a tractable and efficient means to estimate the quantum entanglement present in the system without the need to reconstruct the full $N$-photon state. This effectively local approach not only simplifies the computational process but also delivers a detailed assessment of the entanglement properties. Stabilizer generator expectation values offer these capabilites to serve as benchmarks for these states because by definition, the graph states are composed of a superposition of stabilizer expressions \cite{toth2005entanglement,hein2006entanglement}.
Unlike readily accessible metrics for generation protocol quality, such as rotation fidelities, stabilizer generator expectation values that are directly derived from the $N$-photon quantum state offer an accurate reflection of the system's entanglement characteristics \cite{briegel2001persistent}, even in lossy environments. To construct the stabilizer generator expectation values within the early and late basis specifically for a 1D linear cluster state, we explicitly calculate both second and third order photon correlation functions:
\begin{align}
    \mathcal{G}^{(2)}_{ijkl}(t_1,t_2) &= \langle a^\dagger(t_1)a^\dagger(t_2)a(t_2)a(t_1) \rangle \label{eqn:g2} \\ 
    \mathcal{G}^{(3)}_{ijklmn}(t_1,t_2,t_3) &= \langle a^\dagger(t_1)a^\dagger(t_2)a^\dagger(t_3)a(t_3)a(t_2)a(t_1) \rangle\label{eqn:g3}\,,
\end{align}
where the time arguments $t_i$ are shifted by the time bin length $T$, as indicated by the $E$ (early) and $L$ (late) subsripts of the corresponding correlation functions (cf. Appendix~B). The multi-time correlation functions are calculated by using the von-Neuman equation again on a modified density matrix using the quantum regression theorem \cite{bauch2021ultrafast,glauber1963quantum}. The operator $a^{(\dagger)}$ is the annihilation (creation) operator corresponding to the first ground- and excited state. We refer to the transition between these states as the \textit{cycling transition}.
We then reduce the correlation functions by integrating over all time dimensions such that
\begin{align}
    \overline{\mathcal{G}}^{(2)}_{ijkl} &= \int_0^T\int_{2T}^{3T} dt_1 dt_2 \: \mathcal{G}^{(2)}_{ijkl}(t_1, t_2) ~ \text{and}\\
    \overline{\mathcal{G}}^{(3)}_{ijklmn} &= \int_0^T\int_{2T}^{3T}\int_{4T}^{5T} dt_1 dt_2 dt_3 \: \mathcal{G}^{(3)}_{ijklmn}(t_1, t_2, t_3)\label{eqn:zxz_integrals}\,.
\end{align}
From these photon correlations, we calculate the stabilizer generator expectation values for a 1D linear cluster state:
\begin{align}
    \langle \hat{\Phi}^{(1)} \hat{Z}^{(2)} \rangle &= 2\frac{ \overline{\mathcal{G}}^{(2)}_{EEEL} - \overline{\mathcal{G}}^{(2)}_{ELLL} }{ \overline{\mathcal{G}}^{(2)}_\text{tot} } \label{eqn:xz_gev}
\end{align}
and
\begin{align}
    \langle \hat{Z}^{(1)} \hat{\Phi}^{(2)} \hat{Z}^{(3)}  \rangle
        &= 2\frac{ \overline{\mathcal{G}}^{(3)}_{EEEELE} + \overline{\mathcal{G}}^{(3)}_{LELLLL} }{ \overline{\mathcal{G}}^{(3)}_\text{tot} } \nonumber\\
        &- 2\frac{ \overline{\mathcal{G}}^{(3)}_{EELLLE} + \overline{\mathcal{G}}^{(3)}_{LEEELL} }{ \overline{\mathcal{G}}^{(3)}_\text{tot} }\label{eqn:zxz_gev}\,.
\end{align}
The operator $\hat{\Phi} = \cos(\varphi)\hat{X} + i\sin(\varphi)\hat{Y}$ defines the phase $\varphi$ of the stabilizer generator expectation values which is linearly tunable by adjusting the phase $\vartheta$ of the rotation pulse.
The numerator is normalized with
\begin{align}
    \overline{\mathcal{G}}^{(2)}_\text{tot} = \sum_{i,j \in \{ E, L \} } \overline{\mathcal{G}}^{(2)}_{ijji}
\end{align}
and
\begin{align}
    \overline{\mathcal{G}}^{(3)}_\text{tot} = \sum_{i, j, k \in \{ E, L \} } \overline{\mathcal{G}}^{(3)}_{ijkkji}\,.
\end{align}
\revision{The Pauli spin matrices $\hat{X}$,$\hat{Y}$ and $\hat{Z}$ are multiplied using the Kronecker product.} These expressions are specific to 1D linear cluster states, grow linearly with the temporal photon entanglement \cite{nutz2017proposal}, and can be used to calculate a lower bound for the generated \revision{degree of} entanglement. In the ideal case, the stabilizer generator expectation values approach unity. The necessary configuration for the local bins to calculate any of these stabilizer generator expectation values adheres to the sequence
\begin{align}
    t_0 < t_1 < t_1+T < 2T < t_2 < t_2+T < 4T < \ldots \,,
\end{align}
where $t_0$ marks the commencement of the first bin associated with the expression, and its positioning is flexible within the time frame of the simulation to calculate later generator expectation values. \revision{For $t_0 > 0$, we \revision{shift} the starting time of all of the integrals in \cref{eqn:zxz_integrals}}. Further details are given in Appendix \ref{app:correlations}.

The expectation values $\langle \hat{X}^{(1)} \hat{Z}^{(2)} \rangle$ and $\langle \hat{Y}^{(1)} \hat{Z}^{(2)} \rangle$, as well as $\langle \hat{Z}^{(1)} \hat{X}^{(2)} \hat{Z}^{(3)} \rangle$ and $\langle \hat{Z}^{(1)} \hat{Y}^{(2)} \hat{Z}^{(3)} \rangle$ can be obtained by taking the real or imaginary part in \cref{eqn:xz_gev,eqn:zxz_gev}, respectively. This corresponds to setting $\varphi=0$ ($\varphi=\pi/2$) in the $\Phi$ operator. Later generator expectation values can be calculated by shifting $t_1$ appropriately, indicated by the superscript indices. Note that to classify other graph or cluster states, one may need to evaluate different stabilizer expressions \cite{toth2005entanglement}. Using optimal detunings for $\pi$ and $\pi/2$ rotations, we carry out explicit simulations of the scheme outlined in Section \ref{sec:timebinphotons} as illustrated in \cref{fig:temp_system} b). We adjust the relative phases $\Theta_i$ of the pulses that interact with each ground state, effectively tuning the rotation angle $\vartheta$ in the x-y plane. The phase dependency observed from these adjustments is detailed in \cref{fig:phase_sweep}. Importantly, the magnitude of $\langle \hat{\Phi}^{(1)} \hat{Z}^{(2)} \rangle$ and $\langle \hat{Z}^{(1)} \hat{\Phi}^{(2)} \hat{Z}^{(3)}  \rangle$ consistently remains close to unity, with neither value falling below 0.95 for either property. This suggests that the chosen phase of the rotation pulses predominantly influences the phase of the stabilizer generator expectation values rather than their magnitudes. Furthermore, it is important to recognize that the magnitudes approach unity exclusively under conditions of ideal pulses and for a lossless system. In scenarios deviating from these ideals, the magnitudes can drop significantly below one. Based on the data presented in \cref{fig:phase_sweep}, we deduce that it is feasible to adjust the rotation angle to optimize either the real or the imaginary component of the associated stabilizer generator expectation values, without compromising the overall magnitude of these values, as long as all rotations to generate the cluster state occur in the same plane. In essence, it is essential for all rotation pulses to be phase locked, though the specific phase value itself is inconsequential.

\begin{figure}[!t]
    \centering
    \includegraphics[width=1\linewidth]{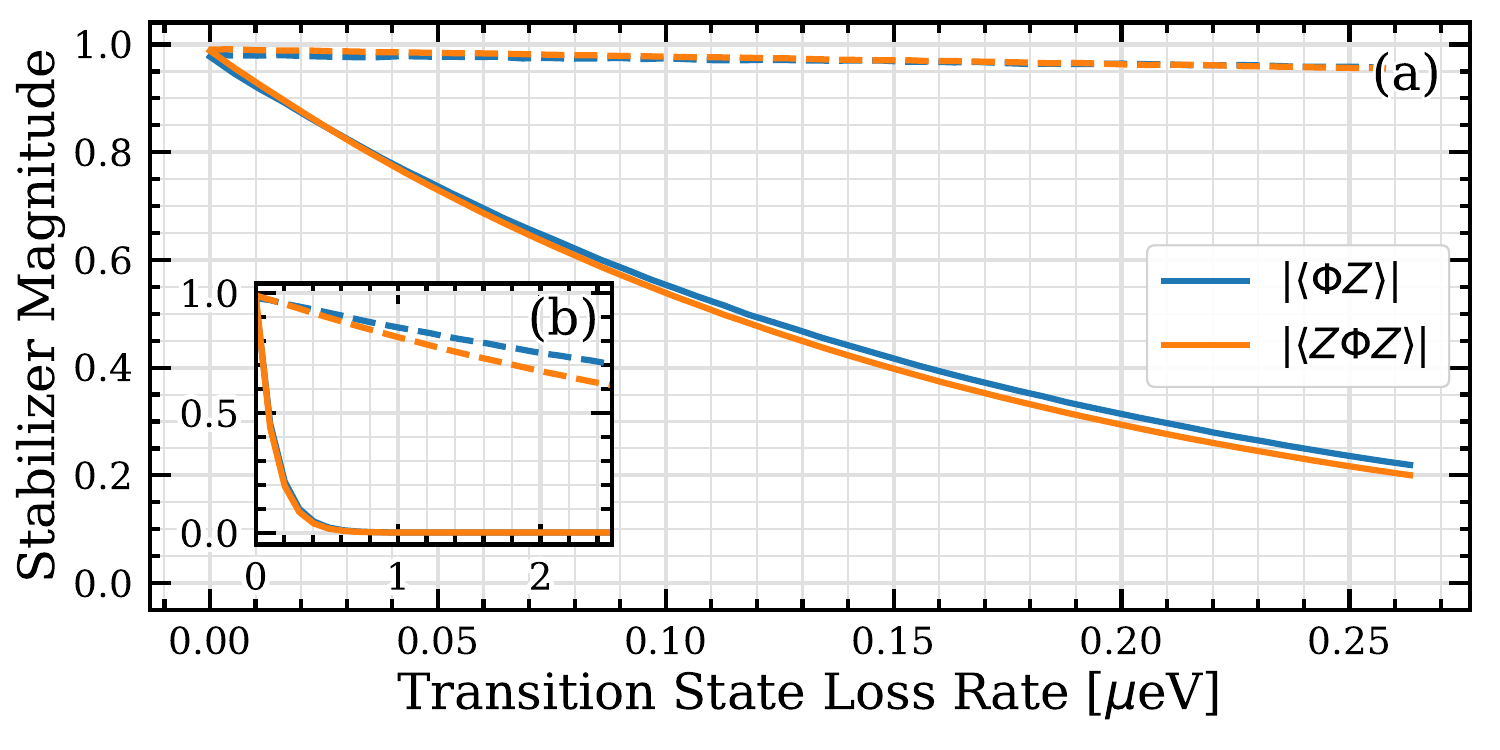}
    \caption{\textbf{Loss dependence of the stabilizer generator expectation values}. Panel \textbf{a)} depicts the magnitudes of the first two stabilizer generator expectation values $|\langle \Phi Z\rangle|$ and $|\langle Z\Phi Z\rangle|$ for different decays of target and unwanted states (dashed lines) and spin dephasing of ground states (solid line). Small decay rates appear to not lower the stabilizer magnitudes significantly, while spin dephasing, and thus the lifetime of the ground state spins, quickly deteriorates the stabilizers. The inset \textbf{b)} shows results for exaggerated loss rates.}
    \label{fig:loss_scan}
\end{figure}

We now explore how rotation fidelities directly influence the quantum characteristics of the generated $N$-photon states, specified by the stabilizer generator expectation values. Given that it is always possible to identify a pulse phase optimizing both characteristics simultaneously, we select an arbitrary phase for our analysis, focusing solely on the magnitude of these properties. Initially, we introduce additional radiative loss exclusively to the transition states, previously set to zero, and adjust it from zero to $2.5 \upmu$eV. The radiative loss rate for the cycling transition is kept constant. In \cref{fig:loss_scan}, we observe that increased radiative losses in the transition state correspond to a direct decrease in the magnitudes of the stabilizers, exhibiting an almost linear relationship for the losses shown. Subsequently, we apply spin dephasing to the ground state coherences, again with rates ranging from zero to $2.5 \upmu$eV. We note that these values lead to dephasing times significantly exceeding the typical achievable values for spin qubits \cite{ruskuc2022nuclear} and provide a solid estimation for the lower bound for the achievable entanglement. In this scenario, we notice a rapid exponential decline in the stabilizer values even for relatively small dephasing rates. It is important to note that this detrimental effect of the ground state dephasing is not apparent in the temporal dynamics of the system, which for small loss values closely resembles the zero-dephasing behavior depicted in \cref{fig:temp_system} b). It is only for losses exceeding $3\upmu$eV that a noticeable difference also becomes visible for the system dynamics.

\begin{figure}[!t]
    \centering
    \includegraphics[width=1\linewidth]{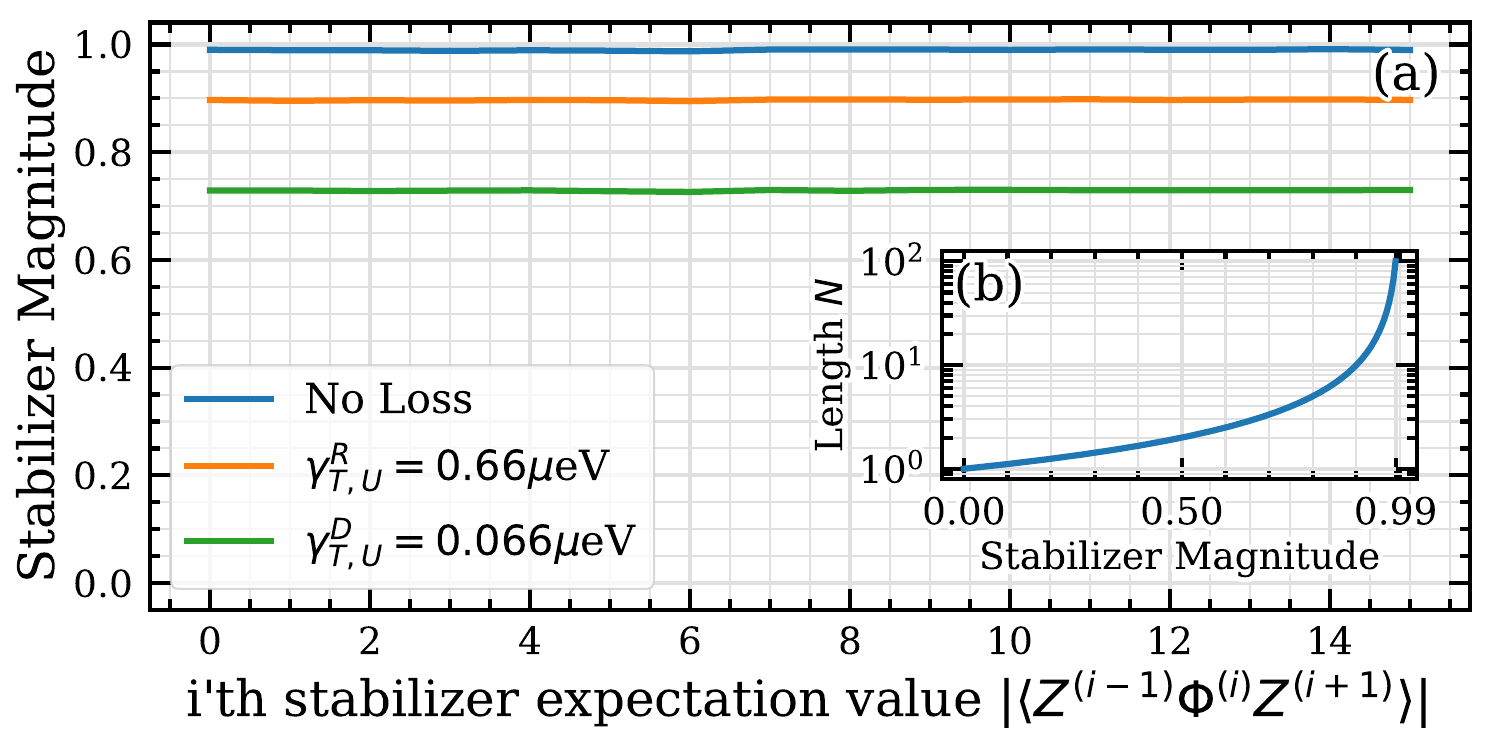}
    \caption{\textbf{Local stabilizer generator expectation values and cluster-state length limitation.} \textbf{a)} Magnitudes of the first 15 $\langle Z\Phi Z\rangle$ expectation values. The lossless case (blue line) is compared to either significant decay of the target and unwanted states (orange line) with $\gamma^R_\text{T,U} = 0.5\cdot\gamma_\text{Cyc} \equiv 1$GHz, which corresponds to a lifetime of the transition states of $1$ns, or moderate spin dephasing of the ground states (green line) with $\gamma^D_\text{T,U} = 0.0375\cdot\gamma_\text{Cyc} \equiv 75$MHz, which corresponds to a spin lifetime greater than $13$ns. \textbf{b)} Lower bound for the possible length of the linear cluster state according to \cref{eqn:length}, assuming all stabilizer expectation values are equal. 
    }
    \label{fig:higher_order}
\end{figure}

It becomes evident that, although the rotation fidelity diminishes somewhat with increasing losses, the nonlinear decay of quantum correlations with increasing spin dephasing does only emerge in a detailed analysis of the quantum \revision{state properties through} stabilizer generators. \revision{Note that we} find that the selection of an optimal bin length is also influenced by the system's loss parameters (not shown), with systems experiencing higher levels of loss potentially benefiting from shorter time bins, which helps to mitigate the adverse effects of dephasing over extended periods of time, optimizing the overall performance of the system under such conditions.

In \cref{fig:higher_order}, we calculate stabilizer generators for up to the 15th expression. Notably, these stabilizers maintain a consistent behavior across a spectrum of loss conditions, meaning all stabilizers remain close to their mean value. \revision{In other words, the standard deviation of all of the stabilizers of a given linear cluster state is small.} This consistency is crucial, as it demonstrates the potential for creating large cluster states; diminishing effects from losses do not appear to accumulate through the protocol cycle. \revision{The stabilizers remain constant because the loss between the first and second $Z$ projections within each stabilizer expression is consistent, making the starting time of the stabilizer evaluation insignificant}.
This stability plays a critical role in the practicality of developing complex cluster and graph states, even within lossy environments. Additionally, it proves advantageous for the numerical analysis that only the first three stabilizer generator expectation values need to be assessed, since \revision{due to their local nature} the subsequent values resemble the initial ones almost exactly.

We employ the witness formalism detailed in \cite{toth2005entanglement} to estimate the lower bound on the number of qubits showing useful entanglement, and consequently, the achievable length of the cluster state \cite{lu2007experimental}. This approach invokes the witness operator $\mathcal{W}$, which for 1D linear cluster states is defined as:
\begin{align}
    \langle\mathcal{W}\rangle = (N-1) - \langle \hat{X}^{(1)} \hat{Z}^{(2)}\rangle - \sum_{i>1} \langle \hat{Z}^{(i-1)} \hat{X}^{(i)} \hat{Z}^{(i+1)}\rangle \label{eqn:witness}\,.
\end{align}
Here, we use pulse configurations that maximize the stabilizer generator expectation values for $\Phi = X$. Evaluating \cref{eqn:witness}, we determine conditions for entanglement by asserting that $\langle\mathcal{W}\rangle < 0$. Assuming uniformity among all stabilizer generator expectation values - with the exception of the initial $\langle X Z\rangle$ - we deduce an expression for the lower bound of the length $N$ of the linear cluster state. We omit the superscript indices from the expressions under the assumption that the expectation values are identical for all indices, which results in
\begin{align}
    N < \frac{\langle \hat{X} \hat{Z} \rangle}{1-\langle \hat{Z} \hat{X} \hat{Z}\rangle} + 1 \label{eqn:length}\,.
\end{align}
Applying \cref{eqn:length} to the three cases presented in \cref{fig:higher_order}, we round down to the nearest integer, resulting in the following length estimates for linear cluster states that can be generated with the investigated scheme: $N_\text{no loss} < 98$, indicating the potential size without any loss mechanisms \revision{(here only limited by the finite pulse lengths and corresponding bandwiths)}; $N_\text{Radiative} < 9$, reflecting the impact of radiative losses; and $N_\text{Dephasing} < 3$, showcasing the limitations imposed by spin dephasing of the ground states, which needs to be minimized for the generation of larger cluster states. We display the \revision{minimum} achievable length of the linear cluster state for a given stabilizer magnitude according to \cref{eqn:length} in \cref{fig:higher_order}.
These results again support that slight enhancements (reductions) in rotation fidelity lead to disproportionately large extensions (decline) in the achievable lengths of cluster states, a phenomenon only perceptible \revision{in a detailed analysis of quantum correlations in the generated multi-partite photon state, here performed} through the lens of stabilizer generator expectation values. While minor percentage improvements in rotation fidelity might not visibly enhance the protocol, our analysis of the stabilizers reveals their profound impact. 

\revision{Although not visualized in this work, our simulations reveal the indistinguishability of the single photons generated in the system to be inherently high, a direct consequence of the photon emission originating from a modified two-level system \cite{lu2021quantum,gschrey2015highly}. Consequently, the indistinguishability is lowered for non-zero spin dephasing of the ground states, again closely mirroring an isolated two level system.} This resemblance of a two-level system ensures that each emitted photon possesses uniform phase and spectral characteristics, critical factors that enhance their indistinguishability \cite{coste2023high}. In quantum information processing, the ability to produce highly indistinguishable photons is paramount, as it facilitates efficient quantum interference, a cornerstone for quantum computing algorithms and photonic quantum information protocols. When dealing with indistinguishable photons, it is also possible to expand the foundational cluster state into multidimensional states through specific measurements \cite{raussendorf2001one}. \revision{For the case with strong radiative losses and dephasing in \cref{fig:higher_order}, the indistinguishability and single-photon purity of the qubit states each stay above $95\%$, approaching unity values when losses approach zero.}

Looking ahead, an intriguing direction for further enhancement of the photonic qubit generation scheme presented would involve the incorporation of optical cavities for the cycling transition. While this approach has not yet been implemented in our current study, it would significantly reduce the lifetime of the excited states \cite{bauch2023demand}. This would increase repetition frequency of photon emission leading to increased source brightness and reducing the required time-bin length, mitigating the undesired influence of spin dephasing of the ground states. This could be instrumental in achieving scalability of the investigated scheme.

\section{Conclusion}
\revision{We conducted an in-depth theoretical study of the generation of time-bin entangled photons using a $\Lambda$-type electronic system, relevant for various physical realizations of emitters. Specifically, we focused our research on a system present in quantum dot molecule composed of two individual quantum dot emitters, resulting in a deterministic source.} Our microscopic numerical investigation reveals that conventional metrics used to analyze excitation-emission protocols, such as fidelities of quantum-state rotations \cite{tiurev2021fidelity}, fall short of fully assessing the usefulness of the correlated quantum states generated, especially when faced with imperfections and losses in a realistic setting. Even when high rotation fidelities are achieved, quantum correlations generated can be surprisingly low, diminishing the usefulness of the generated quantum state. In the present study we use expectation values of stabilizer generators as a more robust measure of correlations in a multi-photon quantum state. These include full sensitivity to dephasing and losses, providing a more nuanced understanding of the system's quantum dynamics \revision{and usefulness as a quantum resource}. For the generation of linear photonic cluster states,  we find that even minor rates of spin dephasing (that hardly affect protocol fidelities) can result in a very significant reduction of stabilizer generator magnitudes and, with it, reduction of quantum correlations.

\revision{Combining these insights we explicitly numerically demonstrate the generation of large linear photonic cluster states comprised of a two-digit number of photonic qubits. These results require optimized timing and phases of laser pulses used for excitation and gating. Notably, maximizing the rotation fidelities towards unity values, where precision is crucial, appears as a requirement for the generation of large numbers of entangled qubits}. In addition to high degrees of quantum correlations achieved, \revision{photons emitted in our calculations} also show inherently high degrees of indistinguishability due to the emission resulting from \revision{an effective} two level system. High photon indistinguishability is a property essential for subsequent quantum interference measurements. Further increase of quantum correlations for given loss parameters could potentially be achieved by reducing lengths of time bins for example by use of optical cavities for the cycling transitions, improving scalability of the generation of correlated multi-photon states for quantum information processing purposes. \revision{Furthermore, by calculating even higher order correlation functions, the evaluation of different, more complicated cluster states can be done, albeit with significantly increased numerical effort.}

\begin{acknowledgments}
This work was supported by the Deutsche Forschungsgemeinschaft (German Research Foundation) through the transregional collaborative research center TRR142/3-2022 (231447078, project C09), the Photonic Quantum Computing initiative (PhoQC) of the state ministry (Ministerium für Kultur und Wissenschaft des Landes Nordrhein-Westfalen), and with computing time provided by the Paderborn Center for Parallel Computing, PC$^2$. We greatfully acknowledge fruitful discussions with Stefan Schulz and Gediminas Juska from Tyndall National Institute \revision{and with Zahra Raissi from Paderborn University}.
\end{acknowledgments}

\appendix

\section{\revision{Hamiltonian \& Equation of Motion}}
\revision{
We use the following Hamiltonian to describe the system numerically:
\begin{align}
    H &= H_0 + H_I \\
    H_0 &= \sum_{\substack{o\in\{\text{G}_1,\text{G}_2,\text{X}_1,\text{X}_2,\text{T},\text{U}\}}} E_o |o\rangle\langle o| \\
    H_I &= \sum_{\substack{i\in\{1,2\}\\S\in\{\text{X}_i,\text{T},\text{U}\}}} \Omega_{i,S}(t)|\text{G}_i\rangle\langle S| + \text{h.c.} ~. \label{hamiltonian}
\end{align}
With generally non-zero ground state energies $E_{\text{G}_i}$, excited state energies $E_{\text{X}_i}$ as well as target and unwanted state energies $E_\text{T}$ and $E_\text{U}$. Here, $\Omega_\text{cyc}(t) = \Omega_{i,\text{X}_i}(t)$, exciting any of the two ground states $\text{G}_i$ into their respective excited states $\text{X}_i$. The target and unwanted transitions are given by $\Omega_{1,2}(t) = \Omega_{i,\text{U}}(t) + \Omega_{i,\text{T}}(t)$, with $i\in\{1,2\}$.
The numerical evaluation is performed using a variable time step Runge Kutta solver, solving the von Neumann equation 
\begin{align}
    \frac{\partial \rho}{\partial t} = -\frac{\i}{\hbar}\left[ H(t), \rho(t) \right] + \mathcal{L}^\text{rad}\left[\rho(t)\right] + \mathcal{L}^\text{dep}\left[\rho(t)\right]
\end{align}
in the interaction frame for $H_0$ with Lindbad rates
\begin{align}
     \mathcal{L}^{\hat{O}}[\rho(t)] = \sum_{\hat{O}}\frac{1}{2}\left(2\hat{O}\rho\hat{O}^\dagger - \hat{O}^\dagger\hat{O}\rho - \rho\hat{O}^\dagger\hat{O}\right)
\end{align}
for the radiative loss ($\hat{O} = \sqrt{\gamma^\text{R}}|i\rangle\langle j|$) and spin dephasing ($\hat{O} = \sqrt{\gamma^\text{D}}\left(|j\rangle\langle j| - |i\rangle\langle i|\right)$ where $E_i < E_j$).
Compare with \cite{bauch2021ultrafast,bauch2023demand} for further details.}

\section{Correlation Functions\label{app:correlations}}

We utilize the well-established second and third order correlation functions, denoted as $\mathcal{G}^{(2)}$ and $\mathcal{G}^{(3)}$, respectively. These functions are essential for understanding the dynamics of time-bin entangled photon pairs within a quantum system. By incorporating the bin length $T$ into our calculations, we adapt these correlation functions to effectively capture the temporal shifts in photon emissions. The evaluation of $\mathcal{G}^{(2)}$ and $\mathcal{G}^{(3)}$ necessitates integration over time variables $t_1$, $t_2$, and additionally $t_3$ for $\mathcal{G}^{(3)}$. The integration intervals are set by the time bin length $T$. 
To ensure accuracy and convergence of our results, we compute these correlation functions over a finely resolved grid, set to $240\times 240 (\times 240)$ grid points for all presented outcomes. This high resolution allows for a detailed examination of the quantum correlations present within our system.

\subsection{General Second and Third Order Correlation Functions}
We introduce the notation for early ($E$) and late ($L$) time bin operators, represented as
\begin{align}
\text{early} = E &\equiv\hat{a}(t) ~\text{ and} \\
\text{late} = L &\equiv\hat{a}(t+T) ~,
\end{align}
respectively.

From the second order correlation functions $\mathcal{G}^{(2)}(t_1,t_2)$ in \cref{eqn:g2} we can construct the two photon density matrix as 
\begin{align}
    \begin{bmatrix}
    \tpm{EE}{EE} & \tpm{EE}{EL} &  \tpm{EE}{LE} & \tpm{EE}{LL} \\
    \tpm{EL}{EE} & \tpm{EL}{EL} &  \tpm{EL}{LE} & \tpm{EL}{LL} \\
    \tpm{LE}{EE} & \tpm{LE}{EL} &  \tpm{LE}{LE} & \tpm{LE}{LL} \\
    \tpm{LL}{EE} & \tpm{LL}{EL} &  \tpm{LL}{LE} & \tpm{LL}{LL}
    \end{bmatrix}
\end{align}
using $\left\{EE,EL,LE,LL\right\}$ as the basis. \revision{In this case, the index ordering corresponds to the operator ordering.}
Analogously, we define the three photon density matrix using $\mathcal{G}^{(3)}(t_1,t_2,t_3)$ in \cref{eqn:g3}, which corresponds to the basis using all of the available three-way permutations of $E$ and $L$. While these would, in theory, suffice for the evaluation of two and three photon entanglement, scalability becomes an obvious problem for larger photon numbers.

\subsection{Examples of Correlation Functions}
For illustrative purposes, we provide examples for computing specific correlation functions. 

For instance, when calculating second order correlations for the operator combination $ELLL$, as required by \cref{eqn:xz_gev}, we adjust all but the first operator in the correlation function using the time bin length $T$. We find
\begin{align}
    \mathcal{G}^{(2)}_{ELLL}(t_1,t_2) &= \langle a^\dagger(t_1)a^\dagger(t_2+T)a(t_2+T)a(t_1+T)\rangle ~.
\end{align}
We then apply the unitary time evolution, accounting for the time ordering operator for each transformation. We use the quantum regression theorem and split the unitary time evolution using $\hat{U}^\dagger(t_i,t_j+T) = \hat{U}^\dagger(t_i,t_j)\hat{U}^\dagger(t_j,t_j+T)$. Furthermore, we denote that any operator encapsuled by appropriate unitary time evolution $\hat{U}^\dagger(t_i,t_j) \ldots \hat{U}(t_i,t_j)$ is written as $\left[\ldots\right]_{t_i\rightarrow t_j}$ for simplicity.
We find that 
\begin{align}
    &\mathcal{G}^{(2)}_{ELLL}(t_1,t_2) = \nonumber\\ &Tr\left\{ \left[\hat{a}\left[\rho(t_1)\hat{a}^\dagger\right]_{t_1\rightarrow t_1+T}\right]_{t_1+T\rightarrow t_2+T} \hat{a}^\dagger \hat{a} \right\} ~,
\end{align}
which requires two time integrals in total. Similarly, we can also calculate a more complex second order correlation function which is not used for the stabilizer generator expectation values in this work. For example, assume we want to calculate the second order correlation function for the operator combination $ELEL$. Due to the nesting of the unitary time evolution, this correlation function will ultimately result in 
\begin{align}
    &\mathcal{G}^{(2)}_{ELEL}(t_1,t_2) = \nonumber\\ &Tr\left\{ \left[\hat{a} \left[\hat{a} \left[\rho(t_1)\hat{a}^\dagger\right]_{t_1\rightarrow t_1+T}\right]_{t_1+T\rightarrow t_2}\right]_{t_2\rightarrow t_2+T}\hat{a}^\dagger \right\} ~,
\end{align}
which is computationally more demanding as it requires three individual time integrals.

A more complex example would be the third order correlation function for the operator combination $EEEELE$, which is required for the calculation of \cref{eqn:zxz_gev}. The computation goes analogously to the second order correlation functions, which results in
\begin{align}
    &\mathcal{G}^{(3)}_{EEEELE}(t_1,t_2,t_3) = \nonumber\\ &Tr\left\{ \left[\hat{a}\left[\left[\hat{a}\rho(t_1)\hat{a}^\dagger\right]_{t_1\rightarrow t_2}\hat{a}^\dagger\right]_{t_2\rightarrow t_2+T}\right]_{t_2+T\rightarrow t_3}\hat{a}^\dagger \hat{a} \right\} ~,
\end{align}
and requires three individual time integrals. Some third order correlation functions require up to five individual time integrals, four of which are nested. This makes evaluation on large grids quickly unfeasible. The remaining operators are calculated analogously.

The data sets employed for the visualizations presented in this study can be accessed via \cite{zenododataset}, while the code utilized to generate the data is accessible through \cite{Bauch_qdacc_2024}.

\bibliography{bibo}

\end{document}